\documentclass[aps,prl,superscriptaddress,twocolumn,showpacs,amsmath,amssymb,floatfix]{revtex4-1}

\usepackage{graphicx}
\usepackage{xspace}
\newcommand{\rto}{$R{}_\text{2}\text{Ti}_\text{2}\text{O}_\text{7}$\xspace}

\newcommand{\dto}{$\text{Dy}{}_\text{2}\text{Ti}_\text{2}\text{O}_\text{7}$\xspace}
\newcommand{\dtox}{($\text{Dy}{}_{\text{1-x}}\text{Y}_{\text{x}})_{\text{2}}\text{Ti}_\text{2}\text{O}_\text{7}$\xspace}
\newcommand{\yto}{$\text{Y}{}_\text{2}\text{Ti}_\text{2}\text{O}_\text{7}$\xspace}

\newcommand{\ea}{\emph{et al.}\xspace}

\newcommand{\ie}{\emph{i.e.}\xspace}
\newcommand{\eg}{\emph{e.g.}\xspace}

\newcommand{\cmag}{$c_{\text{mag}}$\xspace}
\newcommand{\cph}{$c_{\text{ph}}$\xspace}

\begin{document}

\title{Suppression of Pauling's Residual Entropy in Dilute Spin Ice \dtox}

\author{S.~Scharffe}
\author{O.~Breunig}
\author{V.~Cho}
\author{P.~Laschitzky}
\affiliation{{\protect II.}\ Physikalisches Institut, Universit\"at zu K\"oln, Z\"ulpicher Str.\ 77, 50937 K\"oln, Germany}
\author{M.~Valldor}
\affiliation{{\protect II.}\ Physikalisches Institut, Universit\"at zu K\"oln, Z\"ulpicher Str.\ 77, 50937 K\"oln, Germany}
\affiliation{Max-Planck-Institut f\"{u}r Chemische Physik fester Stoffe, Noethnitzer Str.\ 40, 01187 Dresden, Germany}
\author{J.~F.~Welter}
\author{T.~Lorenz}\email[E-mail: ]{tl@ph2.uni-koeln.de}
\affiliation{{\protect II.}\ Physikalisches Institut, Universit\"at zu K\"oln, Z\"ulpicher Str.\ 77, 50937 K\"oln, Germany}

\begin{abstract}

Around 0.5~K, the entropy of the spin-ice \dto has a plateau-like feature close to Pauling's residual entropy derived originally for water ice, but an unambiguous quantification towards lower temperature is prevented by ultra-slow thermal equilibration. Based on specific heat data of \dtox we analyze the influence of non-magnetic dilution on the low-temperature entropy. With increasing $x$, the ultra-slow thermal equilibration rapidly vanishes, the low-temperature entropy systematically decreases and its temperature dependence strongly increases. These data suggest that a non-degenerate ground state is realized in \dtox for intermediate dilution. This contradicts the expected zero-temperature residual entropy obtained from a generalization of Pauling's theory for dilute spin ice, but is supported by Monte Carlo simulations. 
\end{abstract}

\pacs{75.40.-s, 65.40.gd, 75.50.Lk}

\date{\today}

\maketitle

Spin-ice materials attract lots of attention due to their exotic ground state and anomalous excitations~\cite{Bramwell2001,Snyder2001,Ryzhkin2005,Castelnovo2008,Bramwell2009,
Slobinsky2010,Castelnovo2011,Blundell2012}, which arise from a geometric frustration of the magnetic interactions that prevents long-range magnetic order. Prototype spin-ice materials are the pyrochlores \rto with Dy or Ho as $R^{3+}$ ions, which form a network of corner-sharing tetrahedra. The crystal electric field causes a strong Ising anisotropy with local quantization axes pointing from each corner  of a tetrahedron to its center. Thus, each magnetic moment is restricted to one of the $\left\{ 111 \right\}$ directions and may point only either into or out of the tetrahedron. The energy of antiferromagnetic exchange and dipole-dipole interactions is minimized when two spins point into and the other two point out of each tetrahedron. This '2in/2out' ground state is 6-fold degenerate and fulfills Pauling's ice rule describing the hydrogen displacement in water ice with the residual entropy $S_{\text{P}} = (N_{\rm A}k_{\rm B}/2) \ln(3/2)$~\cite{Pauling1935,Nagle1966}. Excitations are created by single spin flips resulting in pairs of tetrahedra with '3in/1out' and '1in/3out' configurations. As a consequence of the ground-state degeneracy, each pair fractionalizes into two individual excitations that can be described as magnetic (anti-)monopoles propagating independently through the lattice~\cite{Ryzhkin2005,Castelnovo2008,Kadowaki2009,Jaubert2009}. The dynamics of these monopole excitations is subject of intense research~\cite{Bramwell2009,Giblin2011,Kolland2013,Grams2014,Scharffe2014a}.

Experimental evidence for Pauling's residual entropy in spin-ice systems stems from specific heat measurements~\cite{Ramirez1999,Hiroi2003,Higashinaka2003,Morris2009} reporting a  practically temperature-independent entropy $S_{\text{ex}}(T\approx 0.4~{\rm K})\simeq S_{\rm P}$.  More recently, however, extremely slow relaxation phenomena were observed for \dto in low-temperature measurements of, \eg, the magnetization~\cite{Matsuhira2011,Revell2012}, ac susceptibility~\cite{Yaraskavitch2012}, thermal transport~\cite{Kolland2012,Scharffe2014} or the specific heat~\cite{Klemke2011,Kolland2012,Pomaranski2013}. Typically, these phenomena set in below $\approx 0.6$~K and signal strongly increasing timescales for the internal thermal equilibration. Therefore, the specific-heat values obtained by standard relaxation-time techniques are too low and $S_{\text{ex}} (T< 0.5~{\rm K})< S_{\text{P}}$ was reported for thermally equilibrated \dto~\cite{Pomaranski2013}. The origin of this discrepancy remains to be clarified. Another open issue is the influence of non-magnetic dilution on the spin-ice ground state~\cite{Sen2014}. By generalizing Pauling's approximation, a non-monotonic dependence $S_{\text{P}}(x)$ as a function of the dilution content $x$ was predicted~\cite{Ke2007}. This was essentially confirmed by recent Monte Carlo (MC) simulations, but for $x> 0.2$ the numerically obtained entropy $S_{\text{MC}}(x,T<0.7~{\rm K})$ falls below the expected $S_{\text{P}}(x)$~\cite{Lin2014}. A quantitative comparison of $S_{\text{MC}}(x,T)$ to experimental data $S_{\text{ex}}(x,T)$ was not done in Ref.~\onlinecite{Lin2014} due to the experimental difficulties, which partly arise from the slow thermal equilibration but also from the uncertainty in estimating the phononic specific heat.     

In this report, we present a detailed specific-heat study of the dilution series \dtox. We find that the slow thermal equilibration is rapidly suppressed with increasing dilution and vanishes for $x\geq 0.2$. For all $x$, the experimentally derived $S_{\text{ex}}(x,T<0.5~{\rm K})$ is smaller than $S_{\text{P}}(x)$ and the deviation increases with $x$. The lowest-temperature ($T=0.4$~K) MC results also overestimate the magnetic entropy of \dtox, but well match the experimental data at $T=0.7$~K. With increasing dilution, our data reveal a systematic increase of the temperature dependence of the low-temperature entropy such that a zero-temperature extrapolation suggests a complete suppression of the residual entropy or, in other words, a non-degenerate ground state for $x>0.2$.

\begin{figure}
    \includegraphics[width=0.9 \linewidth]{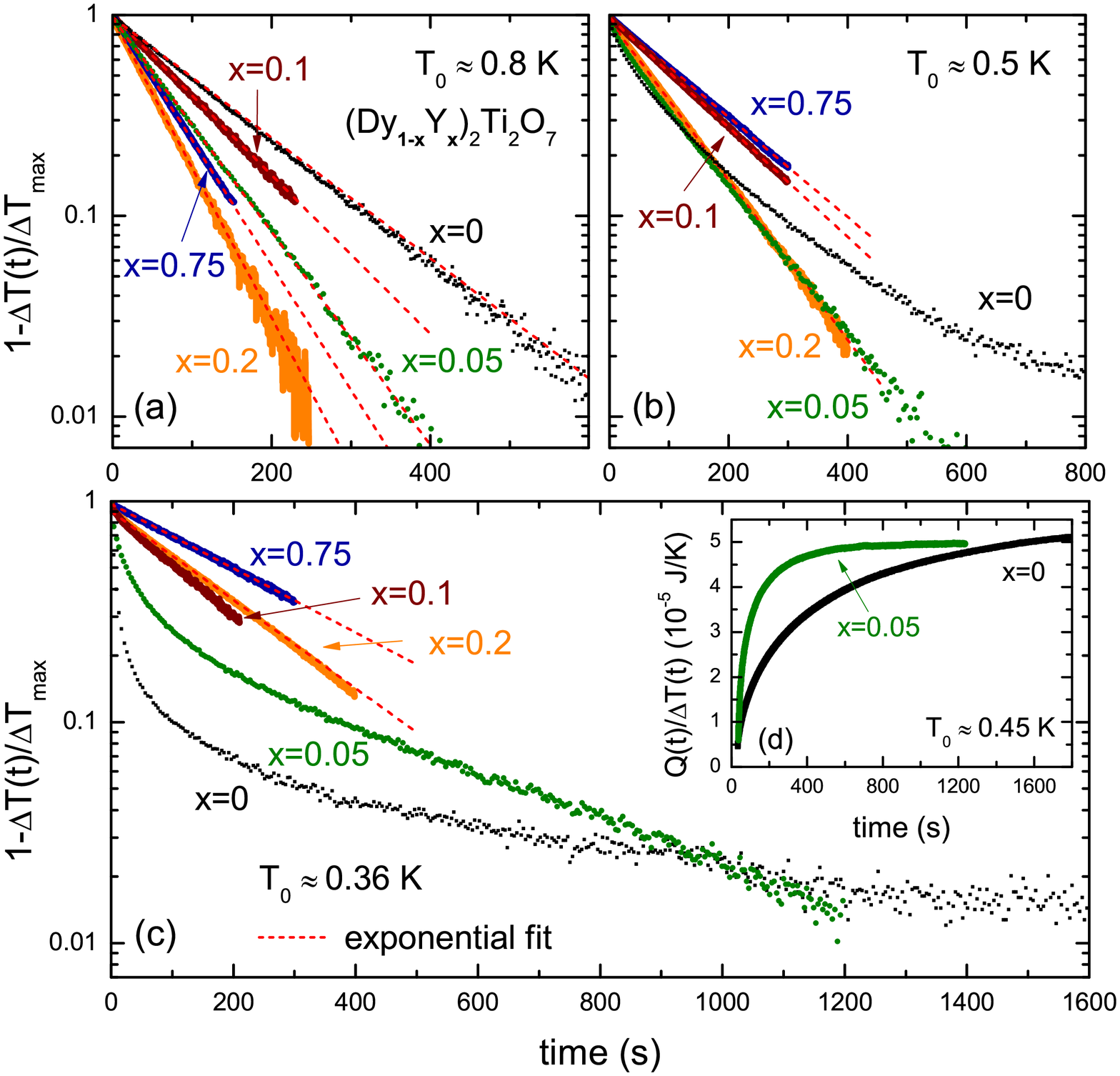}
    \caption{\label{T-gegen-ZeitHFM-komplett}(Color online) Heating curves $ 1-\Delta T(t)/\Delta T_\text{max}$ of \dtox with $x=0$--0.75. Here, $\Delta T(t)=T(t)-T_0$ is measured relative to the base temperature $T_0$ and $\Delta T_{\text{max}}$ is its limiting value. The dashed red lines are fits to those data showing an exponential decay with a single relaxation time. (d) The heat $ Q(t)$ stored in the sample divided by $\Delta T(t)$ approaches the total heat capacity $C$ in the long-time limit.}
  \end{figure}

Oriented \dtox samples of $\approx 20$~mg were cut from large mirror-furnace grown single crystals. The Dy:Y ratio was checked by energy dispersive x-ray diffraction and from the relative decrease of the saturation magnetization. The results of both methods agree within a few percent to the nominal concentration $x$. The specific heat was measured with a home-built calorimeter from about 0.3 to 30~K in magnetic fields of 0, 0.5, and 1~T applied along [100].  In general, the standard relaxation-time method was used, but this method fails if the internal thermal equilibration becomes too slow as is the case in the low-temperature range ($T<0.6$~K) of pure and weakly dilute (see below) spin ice. There, we used a constant heat-flow method analyzing the heating curve over a longer timescale~\cite{Kolland2012}, which is equivalent to the method of Refs.~\onlinecite{tsujii2003,Pomaranski2013} where the specific heat is derived from the temperature-relaxation curve. 
 
\begin{figure}
    \includegraphics[width=0.9 \linewidth]{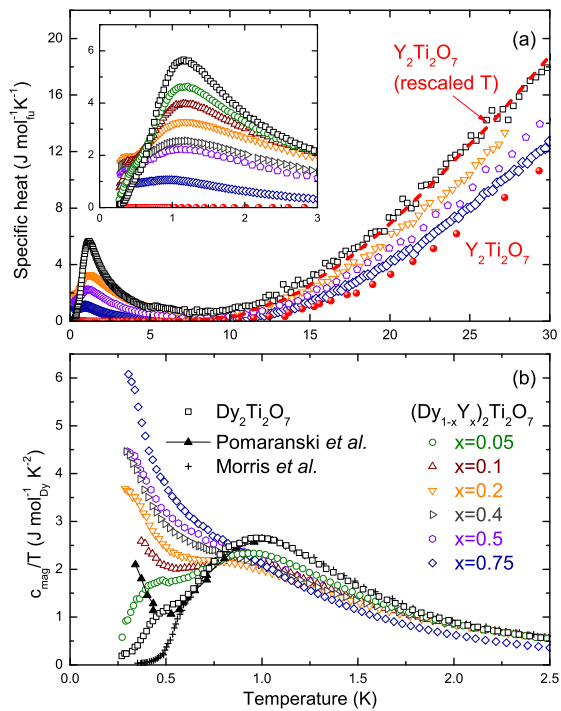}
    \caption{\label{cpT-dot}(Color online) (a) Specific heat $c(T)$ per formula unit (fu) \dtox for selected $x$ with an expanded view of the low-temperature range for all $x$ in the inset. The phononic 
     contributions $c_{\text{ph}}(x,T)$ were estimated by the specific heat of non-magnetic 
    \yto after rescaling the temperature axis such that the data sets match around $T\simeq 25$~K, as is shown for $x=0$ by the dashed line. In (b), the resulting magnetic contribution $c_{\text{mag}}(x,T)=c(x,T)-c_{\text{ph}}(x,T)$ normalized by the Dy content is displayed in the representation $c_{\text{mag}}/T$ versus $T$ for $T\leq 2.5$~K. For \dto, $c_{\text{mag}}/T$ obtained either by a standard relaxation measurement 
    ($+$,\cite{Morris2009}) or  after extremely long-time equilibration ($\blacktriangle$,\cite{Pomaranski2013}) are included.}
  \end{figure}

Fig. \ref{T-gegen-ZeitHFM-komplett} compares typical heating curves of the normalized temperature difference $1- \Delta T(t)/\Delta T_{\text{max}}$ where $\Delta T =T-T_0$ and $T_0$ is the base temperature. At $T_0\simeq 0.8$~K, the heating curves over the entire dilution range in \dtox are straight lines in a semi-logarithmic representation. This is expected if the internal thermal equilibration, where internal means inside the sample as well as between the sample and the platform, is much faster than the thermal relaxation to the external heat bath. The heat capacity is obtained via $C=\tau K$ from the relaxation time $\tau$ and the thermal conductance  $K=P/\Delta T_{\text{max}}$ between the sample platform and the thermal bath with   the heating power $P$. For $T_0\simeq 0.5$~K and 0.36~K, however, the relaxation curves of pure \dto become non-exponential due to slow internal thermal equilibration. In these cases, the heat capacity is obtained from the difference $Q(t)=Pt-\int K \Delta T(t)dt$ between the total dissipated heat and the heat flown from the platform to the bath via $C=Q(t)/\Delta T(t)$, which approaches a constant in the long-time limit, see Fig.~\ref{T-gegen-ZeitHFM-komplett}(d). A weak thermal coupling $K$ is necessary to ensure measurable variations of $\Delta T(t)$ over long-enough times, which restricts the measurements to $\approx 1000$~s in the actual setup. As shown previously~\cite{Kolland2012}, our specific heat data agree well with those of  Ref.~\onlinecite{Klemke2011}, but both data sets are significantly larger than those obtained by the standard relaxation technique~\cite{Morris2009} on \dto for $T<0.6$~K. Our technique and that of Ref.~\onlinecite{Klemke2011} have in common that the heat pulses are analyzed over comparable timescales of up to $\approx 1000$~s. However, according to Ref.~\onlinecite{Pomaranski2013}, the time to reach internal thermal equilibration in \dto drastically increases to several $10^4$s below $\approx 0.4$~K. Such ultra-slow equilibration effects cannot be captured in a setup, whose temperature relaxes significantly faster towards that of the external heat bath. Consequently, our lowest-temperature data (as well as those of Ref.~\onlinecite{Klemke2011}) significantly deviate from the recently published specific heat of \dto that was thermally equilibrated for much longer times~\cite{Pomaranski2013}. As shown in Fig.~\ref{cpT-dot}, our data match those of Ref.~\onlinecite{Pomaranski2013} at $\approx 0.4$~K, but exceed the data obtained by the standard relaxation technique up to $\approx 0.6$~K, above which all data sets finally merge~\cite{nuclear}. 

Concerning dilute \dtox, the low-temperature heating curves for $x=0.05$ and 0.1  also become non-exponential, but this effect is much less pronounced than in pure \dto, and for $x\geq 0.2$ the heating curves remain exponential down to the lowest temperature, see Fig.~\ref{T-gegen-ZeitHFM-komplett}. Thus, our data show that the ultra-slow thermal equilibration in \dto is drastically suppressed by weak non-magnetic dilution. This could result from a suppressed slowing down of the spin-ice dynamics due to an enhanced monopole density in weakly dilute spin ice, because the monopole creation is facilitated close to partially occupied tetrahedra~\cite{Sen2014}.

The specific heat of \dtox measured in zero magnetic field is displayed in Fig.~\ref{cpT-dot}(a). Above $\approx 1$~K, $c(x,T)$ continuously decreases with increasing dilution $x$. Below $\approx  10$~K, this decrease essentially reflects the decreasing amount of magnetic Dy ions, because the magnetic contribution \cmag dominates here. Above 10~K, the phononic contribution \cph starts to dominate and the systematic decrease with increasing $x$ can be traced back to the fact that Y is much lighter than Dy. The molar mass per formula unit (fu) of \dtox decreases from 533 to $386~{\rm g/mol_{fu}}$ between $x=0$ and 1, respectively. Thus, for larger $x$ the eigenfrequencies of the acoustic phonon branches are enhanced and the low-temperature increase of \cph sets in at higher temperature. In order to estimate \cph of the Dy-containing crystals, the temperature axis of the measured $c(T)$ of the non-magnetic \yto was rescaled such that it matches the specific heat of \dtox around 25~K, \ie $c(x=1,r_x\cdot T)=c(x<1,T\simeq 25~{\rm K})$ with scaling factors ranging from $r_x=0.8$ to $0.95$ for $0\leq x\leq 0.75$, respectively. As an example, the resulting \cph of pure \dto is shown as a dashed line in Fig.~\ref{cpT-dot}(a) and the magnetic contributions derived via $c_{\text{mag}}(x,T)= c(x,T)-c(x=1,r_x\cdot T)$ are displayed for all $x\leq 0.75$ in Fig.~\ref{cpT-dot}(b). Here, the data are normalized to the amount of the magnetic Dy ions and are displayed as $c_{\text{mag}}/T$ versus $T$. In this representation, the data for all $x$ almost coincide above 2~K whereas a systematic low-temperature increase of $c_{\text{mag}}/T$ evolves with increasing $x$. The latter observation means that the low-temperature dependence of the entropy $\partial S/\partial T= c_{\text{mag}}/T$ strongly increases with $x$. Note that this conclusion is independent from the uncertainty in estimating the phononic background because any realistic \cph is  negligibly small compared to \cmag in the entire temperature range of Fig.~\ref{cpT-dot}(b). Moreover, it is also essentially independent from the slow thermal equilibration effects, which are only present in the weakly dilute samples at very low temperatures. As can be seen in Fig.~\ref{cpT-dot}(b), the $c_{\text{mag}}/T$ data of thermally equilibrated \dto~\cite{Pomaranski2013} show a low-temperature increase, but still remain below the corresponding data for $x\geq 0.1$.

	  \begin{figure}
    \includegraphics[width=0.9 \linewidth]{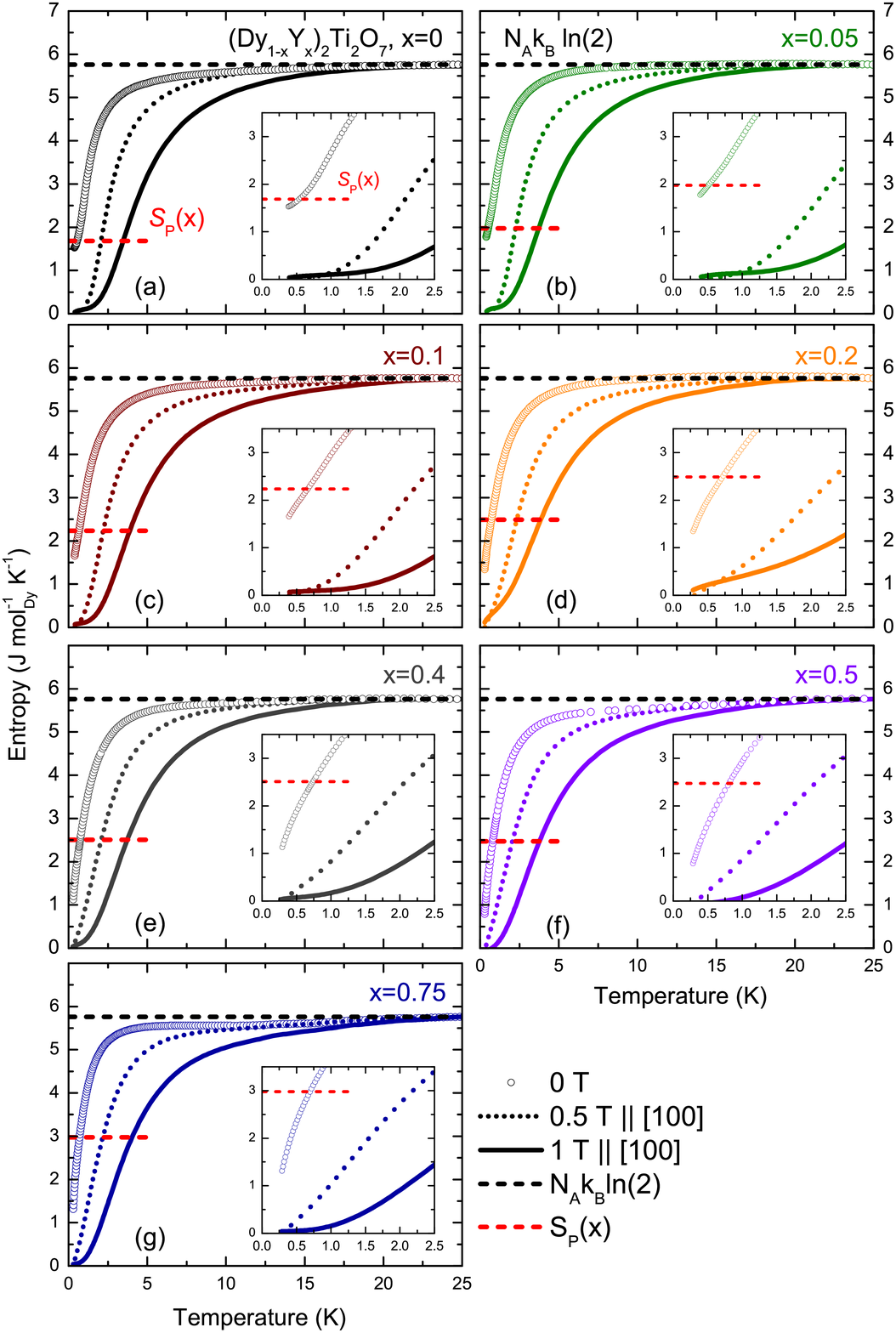}
    \caption{\label{entropy-imfeld}(Color online) Entropy $S_{\text{ex}}(T)$ of \dtox obtained by integration of $c_{\text{mag}}/T$ measured in $B= 0$, 0.5, and 1~T $\| [100]$. All curves are shifted to match the full entropy $S_\infty =N_{\text{A}} k_{\text{B}} \ln(2)$ of a two-level system at $T=25$~K. The red dashed lines mark the residual entropy $S_\text{P}(x)$ expected for $B= 0$ from a generalized Pauling approximation~\cite{Ke2007}.}
  \end{figure}

The magnetic entropy $S(T)$ is obtained by temperature integration of $c_{\text{mag}}/T$ and requires an estimate of \cph. Often \cph is estimated by a Debye model or a simple power law, \eg $\beta T^3 + \beta' T^5$, which match the measured total $c(T)$ around 15~K~\cite{Ramirez1999,Hiroi2003,Ke2007,Lin2014}. However, the corresponding \cmag bears several uncertainties concerning the higher temperature range, as discussed in Ref.~\cite{Lin2014}.  In consequence, the obtained entropy changes for finite magnetic fields do not reach the full entropy $S_\infty =N_{\text{A}} k_{\text{B}} \ln(2)\simeq 5.76$~J/mol\,K, even though the fields are large enough to fully lift the ground-state degeneracy~\cite{Ramirez1999,Hiroi2003,Ke2007}. This can be avoided when \cph is estimated by the measured $c$ of a suitable non-magnetic reference material~\cite{Higashinaka2003}. Here, we use the temperature-rescaled \cph of the non-magnetic \yto and check  the reliability of our procedure by measuring the specific heat of all \dtox samples in $B=0.5$ and 1~T applied along the [100] direction. For this direction, field strengths between 0.5 and 1~T are, on the one hand side, sufficient to reach a fully saturated magnetization at $T\simeq 0.5$~K. On the other hand, such fields are still low enough to reach the full entropy $S_\infty$ of a two-level system around 25~K, where the thermal population of higher-lying crystal field levels is still negligible~\cite{Jana2002,Kitagawa2008}. Figure~\ref{entropy-imfeld} summarizes the magnetic entropy of the series \dtox. In all cases, the integration constants  were adjusted by $S(25~{\rm K})=S_\infty$ and for each composition a magnetic-field independent \cph was used. For all $x$, the low-temperature extrapolations $S(T\rightarrow 0, B\geq 0.5~{\rm T})\rightarrow 0$ and thus clearly confirm the expected vanishing residual entropy in finite fields. 

Now we come to the question, whether there is experimental evidence for a degenerate zero-field ground state in the dilute spin ice \dtox. Because of the ultra-slow thermal equilibration, we restrict this discussion to $T\geq 0.4$~K for the weakly dilute samples with $x\leq 0.1$. As is shown in the inset of Fig.~\ref{entropy-imfeld}(a), the low-temperature entropy of pure \dto is close to the expected $S_\text{P}$ and has a weak, but finite temperature dependence. This approximate plateau-like feature of the entropy is one justification for the description of \dto in terms of a classical spin ice down to these temperatures. The finite slope, which according to Ref.~\cite{Pomaranski2013} further increases below 0.4~K, however, indicates that some kind of ordered ground state ultimately evolves in \dto . One may expect this to occur due to  quantum effects, additional weaker interactions and/or magnetoelastic coupling, but the real ground states of this and other (quantum) spin-ice candidates are in most cases not known, see \eg Refs.~\onlinecite{GingrasMcClarthy2015,RauGingras,Fennell2014PRL}. Recently, various ordered ground states for \dto have been suggested which can arise depending on the strength of quantum tunneling~\cite{McClartyPRB2015}. Moreover, weak non-magnetic dilution may induce transitions to a so-called topological spin glass~\cite{Sen2014}, with a gradual suppression of the residual entropy setting in at $T_c(x)$~\cite{remarkSen2014}.

	  \begin{figure}
    \includegraphics[width=1 \linewidth]{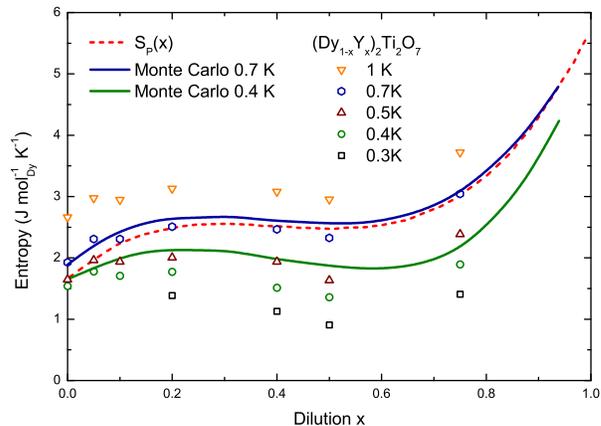}
    \caption{\label{genpauling}(Color online) Comparison of the low-temperature entropy $S_{\text{ex}}(x,T)$ of \dtox 
    (symbols) with the corresponding $S_{\text{MC}}(x,T)$ from MC simulations (solid lines,\cite{Lin2014}) 
    and the zero-temperature residual entropy $S_\text{P}(x)$ (dashed line) expected from a generalized   
    Pauling approximation~\cite{Ke2007}.}
  \end{figure}

Concerning the entropy of the dilute \dtox, the plateau-like feature around 0.5~K rapidly vanishes and the temperature dependence of $S_{\text{ex}}(x,T)$ strongly increases with $x$. Nevertheless, for $x\leq 0.1$ a linear extrapolation $S_{\text{ex}}(x,T\rightarrow 0)$ would still yield a finite zero-temperature residual entropy, what may be interpreted as reminiscence of spin-ice behavior in this intermediate temperature range. Towards larger $x$, however, the slope $\partial S_{\text{ex}}/\partial T= c_{\text{mag}}/T$ becomes so large that an interpretation in terms of a finite residual entropy is no longer justified. For comparison with theoretical predictions, the generalized Pauling residual entropy $S_\text{P}(x)$ from the early work of Ke \ea~\cite{Ke2007} is shown by the dashed lines in Fig.~\ref{entropy-imfeld}. Although $S_\text{P}(x)$ is a zero-temperature result it is significantly larger than the experimental $S_{\text{ex}}(x,T)$, in particular for $x\geq 0.2$. Such a deviation has already been found in a recent comparison of $S_\text{P}(x)$ with the low-temperature entropy $S_{\text{MC}}(x,T)$ obtained by MC simulations~\cite{Lin2014}. In Fig.~\ref{genpauling}, we include $S_{\text{ex}}(x,T)$ of \dtox to this comparison. For $T=0.7$~K, $S_{\text{ex}}(x,T)$ is quantitatively  reproduced by $S_{\text{MC}}(x,T)$ in the entire dilution range $0\leq x \leq 0.75$. In contrast, the lowest-temperature MC data $S_{\text{MC}}(x,T=0.4~{\rm K})$ overestimate the experimental results and essentially reproduce $S_{\text{ex}}(x,T=0.5~{\rm K})$. Thus, an extension of the MC simulations to lower temperatures and including quantum effects would be highly desirable. Concerning the predicted non-monotonic $x$ dependence, a shallow maximum of the entropy around $x \approx 0.2$ is also present in $S_{\text{ex}}(x,T\geq 0.4~{\rm K})$, while below that temperature the slow thermal equilibration for $x\leq 0.1$ prevents a definite conclusion. From $x=0.5$ to $0.75$, the entropy increases again and we think that this reflects the fact that with increasing $x$ the average dipole-dipole interaction decreases. Thus, the temperature-dependent drop of the entropy continuously shifts towards lower temperature and, as a consequence, the entropy at constant temperature continuously increases with $x$.

In conclusion, we find that the ultra-slow thermal equilibration in pure spin ice \dto is rapidly suppressed upon dilution with non-magnetic Y and vanishes completely for $x\geq 0.2$. In general, the low-temperature entropy of \dtox considerably decreases with increasing $x$, whereas its temperature-dependence drastically increases. Thus, there is no experimental evidence for a finite zero-temperature entropy in \dtox above $x \simeq 0.2$, in contrast to the finite $S_\text{P}(x)$ expected from a generalized Pauling approximation~\cite{Ke2007}. Monte Carlo simulations of the low-temperature entropy~\cite{Lin2014}  quantitatively agree with the experimental $S_{\text{ex}}(x,T)$ at $T=0.7$~K, but a systematic deviation develops at lower temperature. Thus, the classical spin-ice model is applicable down to this intermediate temperature range, but additional experimental and theoretical work is necessary to unravel the true ground state of the dilute spin ice \dtox.

\begin{acknowledgments}
We acknowledge financial support by the Deutsche Forschungsgemeinschaft via project LO~818/2-1.
\end{acknowledgments}


\end{document}